\documentclass[prl,twocolumn,amsmath,amssymb]{revtex4}

\usepackage{color}
\usepackage{bm}
\usepackage{dcolumn}
\usepackage[dvips]{graphicx}

\usepackage{ulem}

\begin{document}

\preprint{preprint}

\title{
A Spin Pump Characterized by 
Entanglement Chern Numbers
}

\author{Takahiro Fukui$^1$}
\author{Yasuhiro Hatsugai$^2$}
\affiliation{$^1$Department of Physics, Ibaraki University, Mito 310-8512, Japan}
\affiliation{$^2$Division of Physics, University of Tsukuba, 1-1-1 Tennodai, Tsukuba, Ibaraki 305-8571, Japan}

\date{\today}

\begin{abstract}
  We study a spin pump on a two-leg ladder chain of the Rice--Mele model.
To characterize the spin pump,
we propose the Chern number for the many-body ground state of the entanglement Hamiltonian, which is
referred to as the entanglement Chern number. 
We show that this model has two phases distinguished by the entanglement Chern numbers.
These two phases can be experimentally  verified in cold atoms.

\end{abstract}


\maketitle

Many-body ground states of symmetry-protected topological phases \cite{Schnyder:2008aa,Hasan:2010fk,Qi:2011kx}
are characterized by bulk topological invariants as well as
edge (surface) states. This feature is
known as the bulk-edge correspondence \cite{Halperin:1982uq,Hatsugai:1993fk}.
Two-dimensional systems with no specific symmetries (Class A) \cite{Altland:1997aa,Schnyder:2008aa}
are classified by the Chern number (CN) $c$
\cite{Thouless:1982uq,kohmoto:85}.
Correspondingly, the numbers
of edge states with chirality $\pm$ at a specific boundary, $n_\pm$, are constrained to be $c=n_+-n_-$.
Generically, for a given $c$, a model with $n_+=0$ or $n_-=0$
can be a minimum model.
Although a more generic number of edge states may be possible, 
pair annihilation of edge states with opposite chiralities makes
the minimal realization by chiral edge states most stable.
The symmetry-protected topological phases 
are exceptions to this rule.
For example,
time-reversal (TR) symmetry inducing the Kramers degeneracy allows unusual edge states 
typically with $n_\pm=1$
\cite{Kane:2005ab,Kane:2005aa,Fu:2006aa,Qi:2008aa} even when $c=0$.
These are characterized by the bulk Z$_2$ invariant \cite{Fu:2006aa,Fu:2007aa,Fu:2007fk}.

A weak topological insulator \cite{Fu:2007fk,Moore:2007aa,Roy:2009aa} is another type of nontrivial example.
This material shows two or zero Dirac surface states depending on the surfaces.
Two surface Dirac states appear to be  unstable at first sight
but have stability even against disorder \cite{Ringel:2012xy,Morimoto:2014fk}.
A similar phase has been found also in two dimensions, 
which has a vanishing CN but
shows boundary-dependent edge states \cite{Fukui:2013mz,Yoshimura:2014qf}.
This phase is also referred to as a weak topological phase.
This phase also has stability  since it is 
characterized by 
the entanglement CN (ECN)\cite{Fukui:2014qv,Fukui:2015fk}.

The ECN has been introduced recently as the CN of the many-body ground state of
the entanglement Hamiltonian (EH).
As shown in Ref. \cite{Araki:2016aa}, the ECN can serve as an alternative to the Z$_2$ invariant. Indeed, 
the global phase diagram of the Kane--Mele model \cite{Kane:2005aa} is reproduced by the ECN.
Since the ECN is defined even without TR symmetry, the phase diagram 
in the presence of a magnetic field has also been discussed.
On this basis, in this paper we study a spin pump, 
i.e., a one-dimensional analog of the topological insulator,
which is attracting much current interest 
owing to the recent experimental success of the charge pump \cite{Nakajima:2016aa,Lohse:2016aa}.
We propose a simple spin pump model that is experimentally more feasible than the Fu--Kane model \cite{Fu:2006aa}
at the cost of TR symmetry.

Even without TR symmetry, the ground state of the model we propose has a vanishing CN.
According to the conventional topological classification scheme,
 $c=0$ states with no specific symmetries
are topologically trivial.
As stated in the von Neumann--Wigner theorem, any bulk ground state may be adiabatically connected by the inclusion 
of infinitesimal symmetry-breaking terms. 
However, this is too restrictive when one considers the physical implication of a 
gapped topological system. 
If the mother state is a symmetry-protected topological state, 
daughter states with reduced symmetries still preserve
edge states as the low-energy modes.
Then,  by focusing on the edge states as the key feature of the nontrivial bulk,
  the state is still topologically nontrivial.

For such daughter states, no matter how small symmetry-breaking perturbations are, 
the Z$_2$ invariant is no longer well-defined,
whereas the CN is still zero as long as the perturbations are sufficiently small. 
The ground state of the Kane--Mele model with a weak magnetic field is one example.
Nevertheless,
the ECN suggests a nontrivial ground state even for such a system \cite{Araki:2016aa}.
Thus, we expect that the ECN may be helpful in describing such an almost symmetry-protected topological state.
Unfortunately, there is still a lack of rigorous mathematical foundation for the ECN; thus, we  
need to study more examples other than the Kane--Mele model.
A spin pump is suitable for this purpose since we expect experimental verification in cold atom systems.
Knowing the stability of topological states under the relaxation of the symmetry restriction
is of both theoretical and experimental interest.

To begin with, we review the basic model of a charge pump \cite{Thouless:1983fk}, 
described by the (spinless) Hamiltonian
\cite{Rice:1982qf,Vanderbilt:1993fk,Xiao:2010fk}
$H(t)=\sum_{i,j} c_i^\dagger {\cal H}_{ij}(t) c_{j}$, 
where 
\begin{alignat}1
{\cal H}_{ij}(t)=\frac{t_0+(-)^i\delta(t)}{2}(\delta_{i+1,j}+\delta_{i,j+1})
+(-)^i\Delta(t)\delta_{i,j}
\end{alignat}
with time-dependent parameters
$\delta(t)=\delta_0 \cos (2\pi t/T)$ and $\Delta(t)=\Delta_0\sin (2\pi t/T)$.
Here, we regard $t$ as an external parameter that controls the pumping.
Below, 
this model is referred to as the Rice--Mele model \cite{Rice:1982qf} as in Ref. \cite{Xiao:2010fk}.
The half-filled ground state of this model has 
the CN  $c=1$, implying that
after one period $T$,
all the Wannier states located at the sites adiabatically move together to their neighbors 
\cite{Zak:1989fk,Marzari:1997aa,Fu:2006aa,Thouless:1983fk}.
Recently, a topological charge pump  has been  experimentally observed in ultracold 
atoms 
\cite{Nakajima:2016aa,Lohse:2016aa}.

\begin{figure}[h]
\begin{center}
\includegraphics[width=0.7\linewidth]{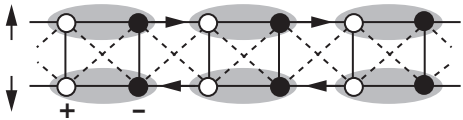}
\caption{
Rice--Mele ladder chain.
Each chain is labeled by the pseudo-spin $\uparrow,\downarrow$, and 
the charges are pumped towards the directions indicated by the horizontal arrows.
Nonequivalent sites in the 
unit cell are labeled by $a=\pm$. 
The vertical solid lines and the dashed lines denote interchain couplings associated with  
a ``magnetic field'' $H_{\rm mg}$ and 
a ``spin-orbit coupling'' $H_{\rm so}$, respectively.
}
\label{f:ladder}
\end{center}
\end{figure}

To study a spin pump, 
let us consider two Rice--Mele chains forming a two-leg ladder.
The time-dependent parameters are defined
such that charges on two chains are pumped towards opposite directions if interchain couplings are
switched off.
Let us distinguish the sites on the two chains by a pseudo-spin $\sigma=\uparrow,\downarrow$,
as illustrated in Fig. \ref{f:ladder}.
The fermion operator is then denoted as $c_{j\sigma}$.
The Hamiltonian is 
\begin{alignat}1
&H(t)=H_0(t)+H_{\rm mf},
\label{LadHam}
\\
&H_0(t)\equiv \sum_{i,j}
\sum_\sigma c_{i\sigma}^\dagger{\cal H}_{ij}(t_\sigma)c_{j\sigma},
\nonumber
\end{alignat}
where 
$t_\uparrow=\theta/2+t$ and $t_\downarrow=\theta/2-t$ with
a relative pumping phase $\theta$, and 
$H_{\rm mf}$ is an interchain coupling defined by
\begin{alignat}1
&H_{\rm mf}=\sum_j \sum_{\sigma,\sigma'}
c_{j\sigma}^\dagger(\bm h\cdot\bm\sigma)_{\sigma\sigma'}c_{j\sigma'}.
\label{Mag}
\end{alignat}
Note that $H_0(t)$ is invariant under TR:
${\cal T}H_0(t){\cal T}^{-1}=H_0(-t)$, where the TR transformation 
is defined by ${\cal T}c_{j}{\cal T}^{-1}=i\sigma_2 Kc_{j}$.
Since $H_{\rm mf}$ breaks TR symmetry, 
it serves as a magnetic field for spins.

If the magnetic field given by Eq. (\ref{Mag}) is replaced by the spin-orbit coupling
\begin{alignat}1
H_{\rm so}=\sum_j \sum_{\sigma,\sigma'}
c_{j\sigma}^\dagger(i\bm e\cdot\bm\sigma)_{\sigma\sigma'}c_{j+1\sigma'}+h.c.,
\end{alignat}
where $\bm e=(e_x,e_y,e_z)$ is the set of real parameters, 
the model simply becomes the Fu-Kane model \cite{Fu:2006aa} for the Z$_2$ spin pump when $\theta=0$.
Even with a finite $\theta$, the model $H_0(t)+H_{\rm so}$ is invariant under TR.
The half-filled ground states of
these models  with $H_{\rm mf}$ and/or $H_{\rm so}$ 
have the vanishing CN $c=0$.

The ground state of the model with the spin-orbit coupling allows two edge states at each boundary.
The level crossing at the TR invariant $t=T/2$ is the Kramers degeneracy, which is 
preserved even with spin-nonconserving terms. 
By the bulk-edge correspondence, 
it is directly related to the Z$_2$ topological invariant of the bulk \cite{Fu:2006aa}.
On the other hand, for 
the model with 
$H_{\rm mf}$,
it is clear that the bulk Z$_2$ invariant cannot be defined.
Nevertheless, the ground state can still be nontrivial, since
as long as
the TR-breaking terms are sufficiently small and 
the bulk gap remains open,
the edge states cannot disappear immediately
even though the Kramers degeneracy is lifted. 
The edge states in this situation again imply the
existence of some nontrivial bulk quantity according to the bulk-edge correspondence.
We expect that the ECN can be used to characterize the bulk.

In addition, the ECN is useful even for a topological insulator with TR symmetry 
\cite{Araki:2016aa}. 
In what follows, we therefore rely on the ECN 
to study the ground state with or without TR symmetry on an equal footing.
To this end, we 
here introduce the notion of the ECN.
The Hamiltonian in Eq. (\ref{LadHam}) is Fourier-transformed to
\begin{alignat}{1}
H(t)=\sum_k\sum_{\alpha,\beta} c_\alpha^\dagger(k) h_{\alpha\beta}(k,t)c_\beta(k),
\label{LadHamM}
\end{alignat}
where $\alpha=\sigma a$ denotes the spin $\sigma$ and bipartite site index $a=\pm$ in the unit cell,
as depicted in Fig. \ref{f:ladder}. 
Let $|G(t)\rangle$ be the half-filled ground state of the Hamiltonian in Eq. (\ref{LadHamM}). 
It is divided into 
each momentum $k$ as $|G(t)\rangle=\prod_k|G(k,t)\rangle$. If the state $|G(k,t)\rangle$ is
decomposed into spin sectors, it generically becomes the sum of the tensor products of the form
\begin{alignat}1
|G(k,t)\rangle=\sum_{i,j} D_{ij}|\Psi_{\uparrow i}(k,t)\rangle\otimes|\Psi_{\downarrow j}(k,t)\rangle ,
\label{SchDec}
\end{alignat}
where $|\Psi_{\sigma i}(k,t)\rangle$ is an orthonormal basis state in the spin-$\sigma$ sector. 
The singular value decomposition of the matrix $D$ leads to the diagonal form
\begin{alignat}1
|G(k,t)\rangle=\sum_i \lambda_i(k,t)|\widetilde\Psi_{\uparrow i}(k,t)\rangle\otimes
|\widetilde\Psi_{\downarrow i}(k,t)\rangle.
\label{SinVal}
\end{alignat}
The normalization of $|G(k,t)\rangle$ requires
$\sum_i\lambda_i^2=1$; thus, 
we can choose $1\ge \lambda_1\ge\lambda_2\ge\cdots\ge0$.
When $\lambda_1\sim1$ and the other $\lambda_i\sim0$ ($i\ge2$), {\it the most dominant tensor-product state}
\begin{alignat}1
|G_{\rm de}(k,t)\rangle\equiv|\widetilde\Psi_{\uparrow 1}(k,t)\rangle\otimes
|\widetilde\Psi_{\downarrow 1}(k,t)\rangle
\label{MosDis}
\end{alignat}
{\it is simply the most disentangled state in $|G(k,t)\rangle$} associated with the spin partition.
The remaining terms associated with $\lambda_i$ ($i\ge2$) are referred to as the {\it residual entanglement} (RE).

Let
$\rho(t)=|G(t)\rangle\langle G(t)|$ be the density matrix of the ground state. Then, it is also 
divided into $\rho(t)=\prod_k\rho(k,t)=\prod_k|G(k,t)\rangle\langle G(k,t)|$.
Tracing out the wavefunctions $|\Psi_{-\sigma i}(k,t)\rangle$ associated with spin $-\sigma$
in Eq. (\ref{SinVal}), 
we obtain the reduced density matrix $\rho_\sigma$
associated with spin $\sigma$
\begin{alignat}1
\rho_\sigma(k,t)&={\rm tr}_{-\sigma} \rho(k,t)
=\sum_i \lambda_i^2
|\widetilde\Psi_{\sigma i}(k,t)\rangle\langle\widetilde\Psi_{\sigma i}(k,t)| .
\label{RedDenMat}
\end{alignat}
It is also parameterized as 
$\rho_\sigma(k,t)\propto \exp[-{\cal H}_\sigma(k,t)]$,
where the EH is defined by 
${\cal H}_\sigma(k,t)=\sum_{a,b}c_{\sigma a}^\dagger(k)h_{\sigma,ab}(k,t)c_{\sigma b}(k)$ 
in the case of noninteracting systems \cite{Peschel:2003uq}.
The entanglement spectrum (ES) is the spectrum of $h_\sigma(k,t)$; 
$\sum_bh_{\sigma,ab}(k,t)\psi_{\sigma, b\mu}(k,t)=\varepsilon_{\sigma,\mu}(k,t)\psi_{\sigma, a\mu}(k,t)$.
To obtain the ES,
it is convenient to utilize the projection operator to the 
ground state defined by
\begin{alignat}1
P_{\beta\alpha}(k,t)
&=\langle G(k,t)|c_{\alpha}^\dagger(k)c_\beta(k)|G(k,t)\rangle
\nonumber\\
&=\sum_{{\rm occupied}~n}\psi_{\beta n}(k,t)\psi_{n\alpha}^\dagger(k,t),
\end{alignat}
where $\psi_{n}(k,t)$ is the wavefunction in the $n$th band of $h(k,t)$ in Eq. (\ref{LadHamM}).
By restricting  $P_{\beta\alpha}$ to 
the spin-$\sigma$ sector such that $\alpha=\sigma a$ and $\beta=\sigma b$, we have the projected 
two-point correlation function $P_{\sigma, ba}(k,t)\equiv P_{\sigma b\sigma a}(k,t)$.
Remarkably, this can alternatively be written \cite{Peschel:2003uq} as
\begin{alignat}1
P_{\sigma, ba}(k,t)&={\rm tr}~c_{\sigma a}^\dagger(k) c_{\sigma b}(k)\rho_\sigma(k,t)
\nonumber\\
&=\sum_\mu \psi_{\sigma, b\mu}(k,t)\frac{1}{e^{\varepsilon_{\sigma, \mu}(k,t)}+1}\psi_{\sigma, a\mu}^\dagger(k,t).
\label{EntSpe}
\end{alignat}
Therefore, an eigenvalue of $P_{\sigma, ba}(k,t)$ is simply the distribution function 
associated with the ES,
and $\psi_{\sigma, \mu}(k,t)$ is a simultaneous eigenfunction of $h_\sigma(k,t)$ and $P_{\sigma}^T(k,t)$.

We now define 
the many-body ground state of the EH $h_\sigma$ as the state with the negative ES
states fully occupied.
This new ground state corresponds to the 
disentangled state $|\widetilde\Psi_{\sigma1}(k,t)\rangle$
in Eq. (\ref{SinVal}). 
Suppose that the ES is gapped. 
The gapped ES implies that the largest $\lambda_1$ term in Eq. (\ref{SinVal}) is unique and that the CN
for the new ground state $|\widetilde\Psi_{\sigma1}(k,t)\rangle$ can be defined.
It is practically computed by the use of the wavefunction
$\psi_{\sigma, \mu}(k,t)$ \cite{Fukui:2014qv,Fukui:2015fk}.
The ECN thus defined will be denoted by $c_\sigma$ below.
This is similar to the spin CN calculated under 
spin-dependent twisted boundary conditions \cite{PhysRevLett.97.036808,Fukui:2007sf}.
However, the computation of the ECN is much simpler 
because it is carried out
in the momentum space using the techniques developed in \cite{FHS05}.
As long as the RE 
is sufficiently small, $c=c_\uparrow+c_\downarrow$
is expected.

The qualitative behavior of 
the model Eq. (\ref{LadHam}) with $H_{\rm mf}$ is as follows.
For simplicity, we restrict our discussion to the cases $\bm h=(h_x,0,0)$.
For a small value of $h_x<h_{c1}$,
the one-particle spectrum of $h(k,t)$ in Eq. (\ref{LadHamM}) has a finite energy gap, 
but it closes
at $h_{c1}$, and when $h_{c2}<h_x$, a new gap opens again.  
In both gapped regions, the CN
is $c=0$, but they are distinguished by 
the ECN, as we show below.

We first consider the small-$h_x$ region. Exactly at $h_x=0$, two spin sectors 
are decoupled; thus,   the pumped charges in one period $T$ are $\Delta q_\uparrow=-\Delta q_\downarrow=1$,
and therefore, the pumped spin $\Delta s_z\equiv(\Delta q_\uparrow-\Delta q_\downarrow)/2=1$ is quantized.
This is ensured by the fact that 
the CNs for the up-spin  and down-spin sectors of the ground state are $1$ and $-1$, respectively.
Once $H_{\rm mf}$ is introduced, 
it is no longer possible to  define such CNs, but
the ECN can be, nevertheless, well-defined.
\cite{Fukui:2014qv,Fukui:2015fk}.
In the region $h_x<h_{c1}$, 
the ES associated with the spin partition is indeed gapped 
and the 
ECNs are computed as
$(c_\uparrow,c_\downarrow)=(1,-1)$.
Therefore, as far as the 
disentangled state 
$|G_{\rm de}\rangle$ in Eq. (\ref{MosDis})
is concerned, topological spin pumping occurs
and the pumped spin is quantized within this disentangled state.
In passing, we mention that for the model with $H_{\rm so}$, the ground state also has the same ECN 
$(1,-1)$.
Thus, in terms of the ECN, both models belong to the same topological class.

\begin{figure}[h]
\begin{center}
\begin{tabular}{cc}
\hspace*{-3mm}
\includegraphics[width=0.50\linewidth]{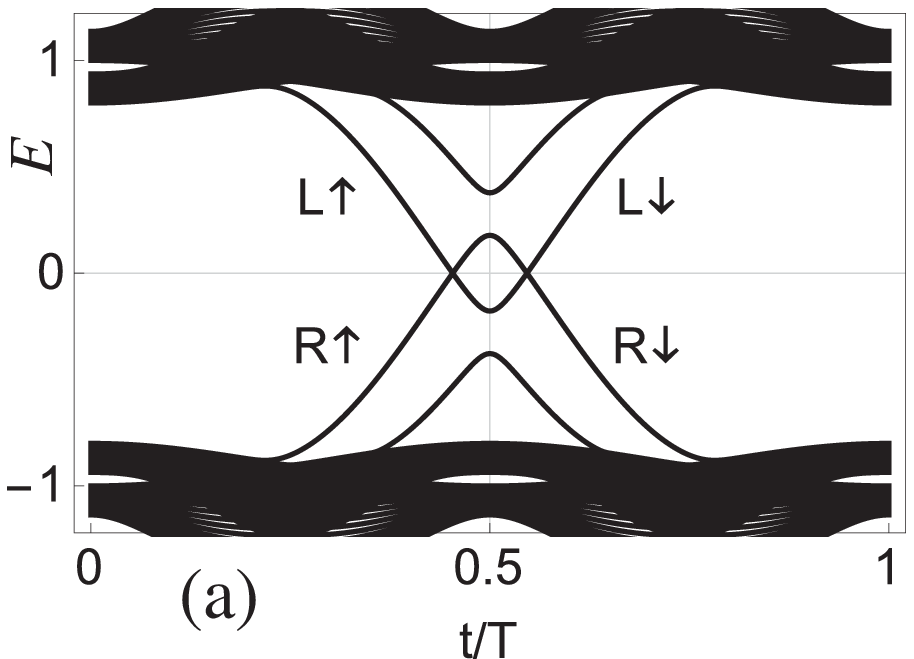}&
\includegraphics[width=0.50\linewidth]{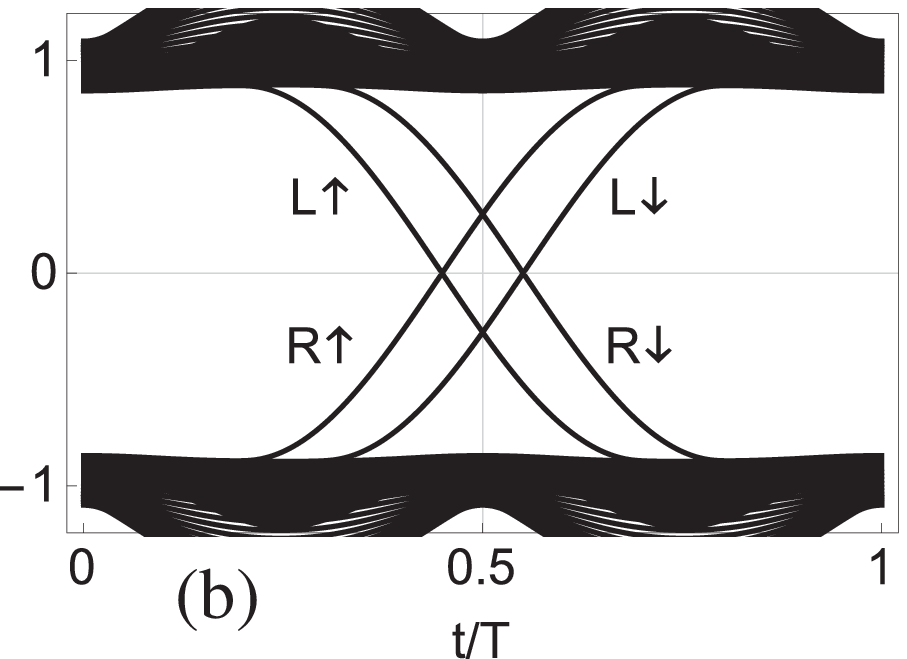}\\
\includegraphics[width=0.50\linewidth]{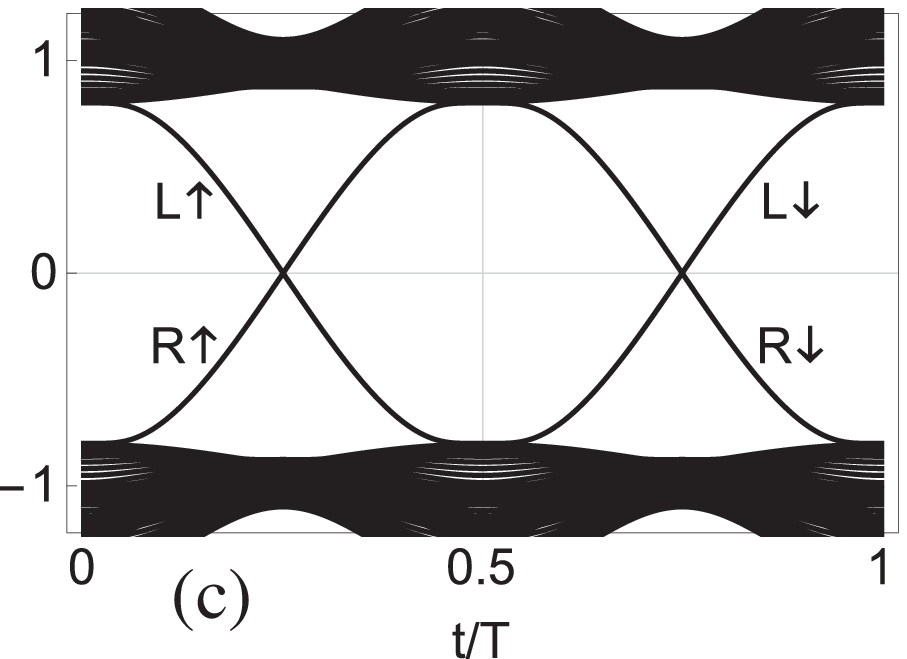}&
\includegraphics[width=0.50\linewidth]{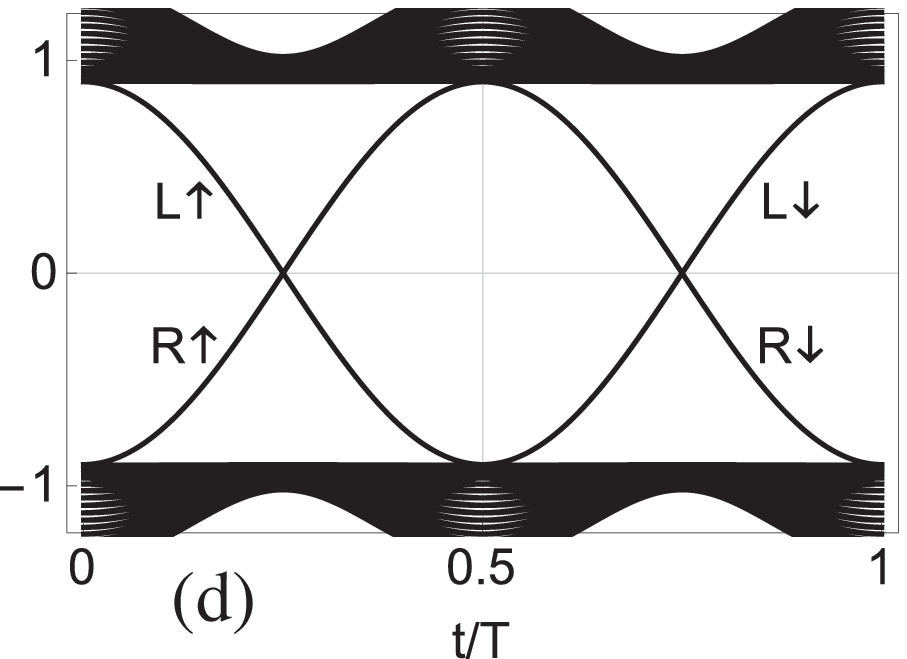}
\end{tabular}
\caption{
Spectra of the Rice--Mele ladder chain:
(a), (c) with $H_{\rm mf}$ ($h_x=0.1$), and (b), (d) 
with $H_{\rm so}$ ($e_x=0.1$).
A finite $\theta$ is introduced: (a), (b) $\theta=\pi/10$, and (c), (d) $\theta=\pi/2$. 
The other parameters are
$t_0=1$, $\delta_0=\Delta_0=0.9$.
``R'' and  ``L'' as well as $\uparrow$ and $\downarrow$ stand for the states localized 
at the right and left ends with spin $\uparrow$ and $\downarrow$, 
respectively.  
}
\label{f:edge}
\end{center}
\end{figure}

However, the physical ground state $|G\rangle$ includes the effect of the RE. 
This is manifest especially in the spectrum of the edge states.
We show in Fig. \ref{f:edge}(a) 
the energy spectrum of a finite Rice--Mele ladder chain with $H_{\rm mf}$. 
We can observe gapped edge states. 
Spectrum (b) is in sharp contrast to spectrum (a),
in which TR invariance guarantees the Kramers' degeneracy at the TR-invariant point $t=T/2$.
 
So far, we have argued that
as long as the symmetry-breaking perturbation is small and the RE is also small,
the dominant component of $|G\rangle$, namely,
the disentangled state $|G_{\rm de}\rangle$, can be topologically nontrivial. Here, 
TR invariance is unimportant.
Correspondingly, the edge states exist in the bulk gap, even though they are gapped 
in the absence of TR symmetry.
When the gapped edge states are due to the nontrivial ECN of the disentangled state, 
it may be possible to make them gapless without closing the bulk gap. 
Indeed, in the present case, if the edge states are shifted in $t$ by changing 
the relative pumping phase $\theta$, spectrum (c) with $H_{\rm mf}$ becomes  
very similar to spectrum (d) with $H_{\rm so}$.
Experimentally, 
it may be possible to observe the pumped charge at each chain and thus the pumped spin.
The observables in the spin-$\sigma$ sector are 
affected by
the RE, and therefore,
the observed pumped spin is not quantized.
However, it should be stressed that the experimental observation of a finite spin pump suggests 
a nontrivial ECN.

\begin{figure}[h]
\begin{center}
\begin{tabular}{cc}
\hspace*{-3mm}
\includegraphics[width=0.50\linewidth]{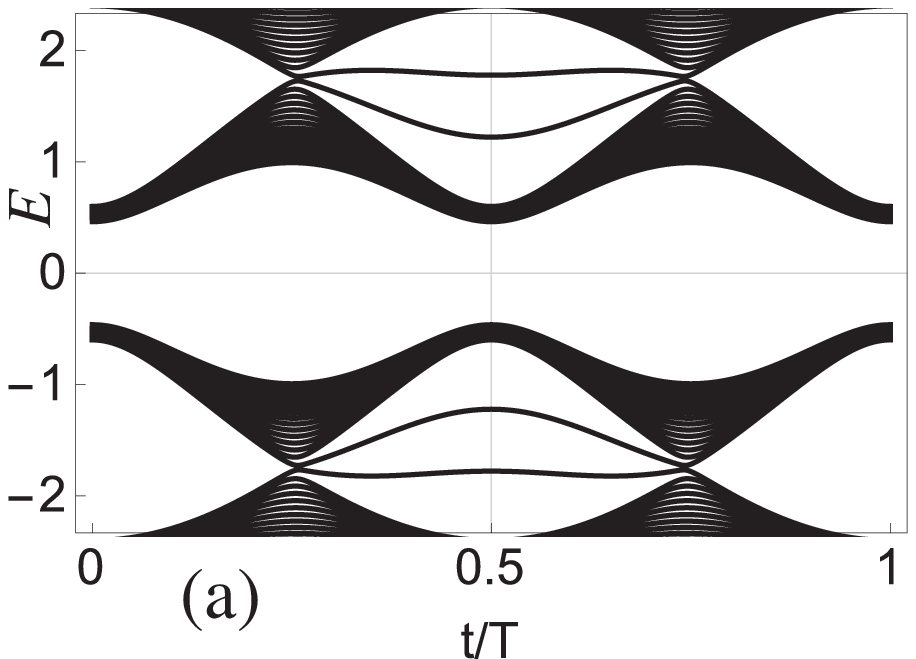}&
\includegraphics[width=0.49\linewidth]{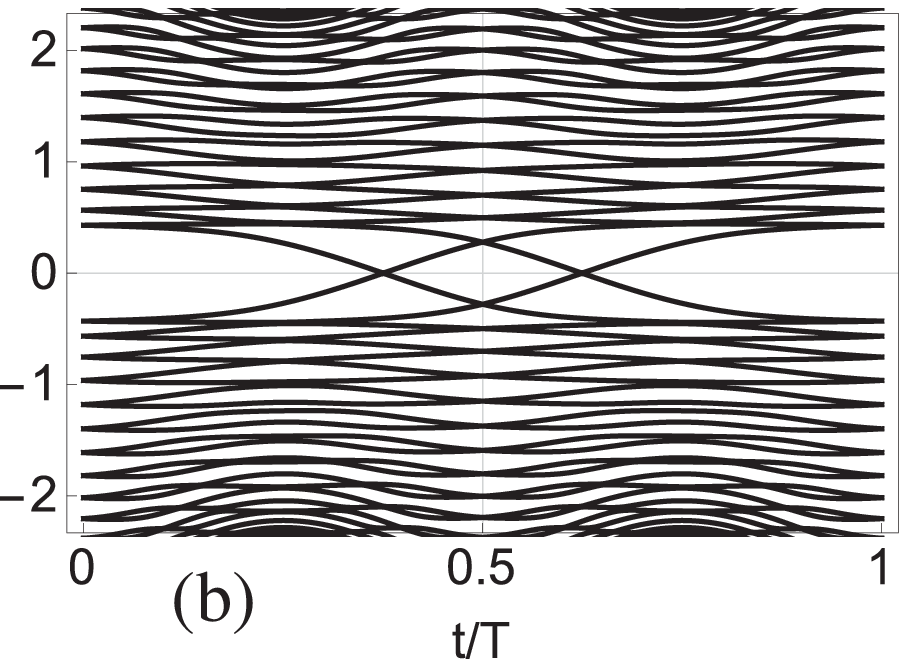}
\end{tabular}
\caption{
Spectra for a larger interchain coupling: (a) $h_x=1.5$ and (b) $e_x=1.5$. 
Other parameters are the same as in Fig. \ref{f:edge}.
}
\label{f:edge_no}
\end{center}
\end{figure}

For a strong magnetic field, $h_x>h_{c2}$,
the ECN associated with the spin is ill-defined, since the spins are strongly entangled and
the ES of $P_{\sigma}$ in (\ref{EntSpe}) becomes gapless.
On the other hand, a new partition associated with $a=\pm$ may be useful, 
since for a large inter-chain coupling, particularly for a large $h_x$,
local interchain dimerized states are expected. 
To see this, we introduce the bipartition with respect to $a=\pm$ and
calculate the 
ES associated with the reduced density matrix $\rho_a$, where $a=\pm$, by 
tracing out $-a$. This can be obtained by the use of the correlation matrix $P_{\beta\alpha}(k,t)$
by restricting $\alpha$ and $\beta$ to $a\sigma$ and $a\sigma'$, which is denoted by
$P_{a,\sigma\sigma'}=P_{a\sigma a\sigma'}$ with $a=\pm$ fixed.
In the region $h>h_{c2}$, it turns out that such a partition gives a gapped ES but the
ECNs are trivial, $(c_+,c_-)=(0,0)$.
Thus, we conclude that the two gapped phases $h_x<h_{c1}$ and $h_{c2}<h_x$
are distinct.
This can also be seen from the spectrum in Fig. \ref{f:edge_no}(a), in which no sign of edge states is observed. 
Contrary to this model, the model with $H_{\rm so}$ exhibits a bulk gap for a rather strong spin-orbit coupling,
as shown in Fig. \ref{f:edge_no}(b). 

In summary, we have introduced the ECN and
proposed a spin pump model that shows distinct phases 
characterized by the ECN.
More detailed analysis based on 
the bulk-edge correspondence in a topological pumping \cite{Wang:2013fk_pump,Hatsugai:2016aa}
is an interesting future issue.
We expect that
the concept of the ECN will give us a way of studying the stability of 
symmetry-protected topological phases against
symmetry-breaking perturbations, which will also open
up the possibility of the experimental realization of 
topological phenomena.

We would like to thank A. Sawada, S. Nakajima, and Y. Takahashi for fruitful discussions.
This work was supported in part by Grants-in-Aid for Scientific Research Numbers 
25400388 (TF), 
16K13845 (YH), and 26247064
from Japan Society for the Promotion of Science.

\bibliography{17204}

\begin{thebibliography}{37}
\expandafter\ifx\csname natexlab\endcsname\relax\def\natexlab#1{#1}\fi
\expandafter\ifx\csname bibnamefont\endcsname\relax
  \def\bibnamefont#1{#1}\fi
\expandafter\ifx\csname bibfnamefont\endcsname\relax
  \def\bibfnamefont#1{#1}\fi
\expandafter\ifx\csname citenamefont\endcsname\relax
  \def\citenamefont#1{#1}\fi
\expandafter\ifx\csname url\endcsname\relax
  \def\url#1{\texttt{#1}}\fi
\expandafter\ifx\csname urlprefix\endcsname\relax\def\urlprefix{URL }\fi
\providecommand{\bibinfo}[2]{#2}
\providecommand{\eprint}[2][]{\url{#2}}

\bibitem[{\citenamefont{Schnyder et~al.}(2008)\citenamefont{Schnyder, Ryu,
  Furusaki, and Ludwig}}]{Schnyder:2008aa}
\bibinfo{author}{\bibfnamefont{A.~P.} \bibnamefont{Schnyder}},
  \bibinfo{author}{\bibfnamefont{S.}~\bibnamefont{Ryu}},
  \bibinfo{author}{\bibfnamefont{A.}~\bibnamefont{Furusaki}}, \bibnamefont{and}
  \bibinfo{author}{\bibfnamefont{A.~W.~W.} \bibnamefont{Ludwig}},
  \bibinfo{journal}{Phys. Rev. B} \textbf{\bibinfo{volume}{78}},
  \bibinfo{pages}{195125} (\bibinfo{year}{2008}).

\bibitem[{\citenamefont{Hasan and Kane}(2010)}]{Hasan:2010fk}
\bibinfo{author}{\bibfnamefont{M.~Z.} \bibnamefont{Hasan}} \bibnamefont{and}
  \bibinfo{author}{\bibfnamefont{C.~L.} \bibnamefont{Kane}},
  \bibinfo{journal}{Rev. Mod. Phys.} \textbf{\bibinfo{volume}{82}},
  \bibinfo{pages}{3045} (\bibinfo{year}{2010}).

\bibitem[{\citenamefont{Qi and Zhang}(2011)}]{Qi:2011kx}
\bibinfo{author}{\bibfnamefont{X.-L.} \bibnamefont{Qi}} \bibnamefont{and}
  \bibinfo{author}{\bibfnamefont{S.-C.} \bibnamefont{Zhang}},
  \bibinfo{journal}{Rev. Mod. Phys.} \textbf{\bibinfo{volume}{83}},
  \bibinfo{pages}{1057} (\bibinfo{year}{2011}).

\bibitem[{\citenamefont{Halperin}(1982)}]{Halperin:1982uq}
\bibinfo{author}{\bibfnamefont{B.~I.} \bibnamefont{Halperin}},
  \bibinfo{journal}{Phys. Rev. B} \textbf{\bibinfo{volume}{25}},
  \bibinfo{pages}{2185} (\bibinfo{year}{1982}).

\bibitem[{\citenamefont{Hatsugai}(1993)}]{Hatsugai:1993fk}
\bibinfo{author}{\bibfnamefont{Y.}~\bibnamefont{Hatsugai}},
  \bibinfo{journal}{Phys. Rev. Lett.} \textbf{\bibinfo{volume}{71}},
  \bibinfo{pages}{3697} (\bibinfo{year}{1993}).

\bibitem[{\citenamefont{Altland and Zirnbauer}(1997)}]{Altland:1997aa}
\bibinfo{author}{\bibfnamefont{A.}~\bibnamefont{Altland}} \bibnamefont{and}
  \bibinfo{author}{\bibfnamefont{M.~R.} \bibnamefont{Zirnbauer}},
  \bibinfo{journal}{Phys. Rev. B} \textbf{\bibinfo{volume}{55}},
  \bibinfo{pages}{1142} (\bibinfo{year}{1997}).

\bibitem[{\citenamefont{Thouless et~al.}(1982)\citenamefont{Thouless, Kohmoto,
  Nightingale, and den Nijs}}]{Thouless:1982uq}
\bibinfo{author}{\bibfnamefont{D.~J.} \bibnamefont{Thouless}},
  \bibinfo{author}{\bibfnamefont{M.}~\bibnamefont{Kohmoto}},
  \bibinfo{author}{\bibfnamefont{M.~P.} \bibnamefont{Nightingale}},
  \bibnamefont{and} \bibinfo{author}{\bibfnamefont{M.}~\bibnamefont{den Nijs}},
  \bibinfo{journal}{Phys. Rev. Lett.} \textbf{\bibinfo{volume}{49}},
  \bibinfo{pages}{405} (\bibinfo{year}{1982}).

\bibitem[{\citenamefont{Kohmoto}(1985)}]{kohmoto:85}
\bibinfo{author}{\bibfnamefont{M.}~\bibnamefont{Kohmoto}},
  \bibinfo{journal}{Annals of Physics} \textbf{\bibinfo{volume}{160}},
  \bibinfo{pages}{343} (\bibinfo{year}{1985}).

\bibitem[{\citenamefont{Kane and Mele}(2005{\natexlab{a}})}]{Kane:2005ab}
\bibinfo{author}{\bibfnamefont{C.~L.} \bibnamefont{Kane}} \bibnamefont{and}
  \bibinfo{author}{\bibfnamefont{E.~J.} \bibnamefont{Mele}},
  \bibinfo{journal}{Phys. Rev. Lett.} \textbf{\bibinfo{volume}{95}},
  \bibinfo{pages}{226801} (\bibinfo{year}{2005}{\natexlab{a}}).

\bibitem[{\citenamefont{Kane and Mele}(2005{\natexlab{b}})}]{Kane:2005aa}
\bibinfo{author}{\bibfnamefont{C.~L.} \bibnamefont{Kane}} \bibnamefont{and}
  \bibinfo{author}{\bibfnamefont{E.~J.} \bibnamefont{Mele}},
  \bibinfo{journal}{Phys. Rev. Lett.} \textbf{\bibinfo{volume}{95}},
  \bibinfo{pages}{146802} (\bibinfo{year}{2005}{\natexlab{b}}).

\bibitem[{\citenamefont{Fu and Kane}(2006)}]{Fu:2006aa}
\bibinfo{author}{\bibfnamefont{L.}~\bibnamefont{Fu}} \bibnamefont{and}
  \bibinfo{author}{\bibfnamefont{C.~L.} \bibnamefont{Kane}},
  \bibinfo{journal}{Phys. Rev. B} \textbf{\bibinfo{volume}{74}},
  \bibinfo{pages}{195312} (\bibinfo{year}{2006}).

\bibitem[{\citenamefont{Qi et~al.}(2008)\citenamefont{Qi, Hughes, and
  Zhang}}]{Qi:2008aa}
\bibinfo{author}{\bibfnamefont{X.-L.} \bibnamefont{Qi}},
  \bibinfo{author}{\bibfnamefont{T.~L.} \bibnamefont{Hughes}},
  \bibnamefont{and} \bibinfo{author}{\bibfnamefont{S.-C.} \bibnamefont{Zhang}},
  \bibinfo{journal}{Phys. Rev. B} \textbf{\bibinfo{volume}{78}},
  \bibinfo{pages}{195424} (\bibinfo{year}{2008}).

\bibitem[{\citenamefont{Fu et~al.}(2007)\citenamefont{Fu, Kane, and
  Mele}}]{Fu:2007aa}
\bibinfo{author}{\bibfnamefont{L.}~\bibnamefont{Fu}},
  \bibinfo{author}{\bibfnamefont{C.~L.} \bibnamefont{Kane}}, \bibnamefont{and}
  \bibinfo{author}{\bibfnamefont{E.~J.} \bibnamefont{Mele}},
  \bibinfo{journal}{Phys. Rev. Lett.} \textbf{\bibinfo{volume}{98}},
  \bibinfo{pages}{106803} (\bibinfo{year}{2007}).

\bibitem[{\citenamefont{Fu and Kane}(2007)}]{Fu:2007fk}
\bibinfo{author}{\bibfnamefont{L.}~\bibnamefont{Fu}} \bibnamefont{and}
  \bibinfo{author}{\bibfnamefont{C.~L.} \bibnamefont{Kane}},
  \bibinfo{journal}{Phys. Rev. B} \textbf{\bibinfo{volume}{76}},
  \bibinfo{pages}{045302} (\bibinfo{year}{2007}).

\bibitem[{\citenamefont{Moore and Balents}(2007)}]{Moore:2007aa}
\bibinfo{author}{\bibfnamefont{J.~E.} \bibnamefont{Moore}} \bibnamefont{and}
  \bibinfo{author}{\bibfnamefont{L.}~\bibnamefont{Balents}},
  \bibinfo{journal}{Phys. Rev. B} \textbf{\bibinfo{volume}{75}},
  \bibinfo{pages}{121306} (\bibinfo{year}{2007}).

\bibitem[{\citenamefont{Roy}(2009)}]{Roy:2009aa}
\bibinfo{author}{\bibfnamefont{R.}~\bibnamefont{Roy}}, \bibinfo{journal}{Phys.
  Rev. B} \textbf{\bibinfo{volume}{79}}, \bibinfo{pages}{195321}
  (\bibinfo{year}{2009}).

\bibitem[{\citenamefont{Ringel et~al.}(2012)\citenamefont{Ringel, Kraus, and
  Stern}}]{Ringel:2012xy}
\bibinfo{author}{\bibfnamefont{Z.}~\bibnamefont{Ringel}},
  \bibinfo{author}{\bibfnamefont{Y.~E.} \bibnamefont{Kraus}}, \bibnamefont{and}
  \bibinfo{author}{\bibfnamefont{A.}~\bibnamefont{Stern}},
  \bibinfo{journal}{Phys. Rev. B} \textbf{\bibinfo{volume}{86}},
  \bibinfo{pages}{045102} (\bibinfo{year}{2012}).

\bibitem[{\citenamefont{Morimoto and Furusaki}(2014)}]{Morimoto:2014fk}
\bibinfo{author}{\bibfnamefont{T.}~\bibnamefont{Morimoto}} \bibnamefont{and}
  \bibinfo{author}{\bibfnamefont{A.}~\bibnamefont{Furusaki}},
  \bibinfo{journal}{Phys. Rev. B} \textbf{\bibinfo{volume}{89}},
  \bibinfo{pages}{035117} (\bibinfo{year}{2014}).

\bibitem[{\citenamefont{Fukui et~al.}(2013)\citenamefont{Fukui, Imura, and
  Hatsugai}}]{Fukui:2013mz}
\bibinfo{author}{\bibfnamefont{T.}~\bibnamefont{Fukui}},
  \bibinfo{author}{\bibfnamefont{K.-I.} \bibnamefont{Imura}}, \bibnamefont{and}
  \bibinfo{author}{\bibfnamefont{Y.}~\bibnamefont{Hatsugai}},
  \bibinfo{journal}{J. Phys. Soc. Jpn.} \textbf{\bibinfo{volume}{82}},
  \bibinfo{pages}{073708} (\bibinfo{year}{2013}).

\bibitem[{\citenamefont{Yoshimura et~al.}(2014)\citenamefont{Yoshimura, Imura,
  Fukui, and Hatsugai}}]{Yoshimura:2014qf}
\bibinfo{author}{\bibfnamefont{Y.}~\bibnamefont{Yoshimura}},
  \bibinfo{author}{\bibfnamefont{K.-I.} \bibnamefont{Imura}},
  \bibinfo{author}{\bibfnamefont{T.}~\bibnamefont{Fukui}}, \bibnamefont{and}
  \bibinfo{author}{\bibfnamefont{Y.}~\bibnamefont{Hatsugai}},
  \bibinfo{journal}{Phys. Rev. B} \textbf{\bibinfo{volume}{90}},
  \bibinfo{pages}{155443} (\bibinfo{year}{2014}).

\bibitem[{\citenamefont{Fukui and Hatsugai}(2014)}]{Fukui:2014qv}
\bibinfo{author}{\bibfnamefont{T.}~\bibnamefont{Fukui}} \bibnamefont{and}
  \bibinfo{author}{\bibfnamefont{Y.}~\bibnamefont{Hatsugai}},
  \bibinfo{journal}{J. Phys. Soc. Jpn.} \textbf{\bibinfo{volume}{83}},
  \bibinfo{pages}{113705} (\bibinfo{year}{2014}).

\bibitem[{\citenamefont{Fukui and Hatsugai}(2015)}]{Fukui:2015fk}
\bibinfo{author}{\bibfnamefont{T.}~\bibnamefont{Fukui}} \bibnamefont{and}
  \bibinfo{author}{\bibfnamefont{Y.}~\bibnamefont{Hatsugai}},
  \bibinfo{journal}{J. Phys. Soc. Jpn.} \textbf{\bibinfo{volume}{84}},
  \bibinfo{pages}{043703} (\bibinfo{year}{2015}).

\bibitem[{\citenamefont{Araki et~al.}(2016)\citenamefont{Araki, Kariyado,
  Fukui, and Hatsugai}}]{Araki:2016aa}
\bibinfo{author}{\bibfnamefont{H.}~\bibnamefont{Araki}},
  \bibinfo{author}{\bibfnamefont{T.}~\bibnamefont{Kariyado}},
  \bibinfo{author}{\bibfnamefont{T.}~\bibnamefont{Fukui}}, \bibnamefont{and}
  \bibinfo{author}{\bibfnamefont{Y.}~\bibnamefont{Hatsugai}},
  \bibinfo{journal}{J. Phys. Soc. Jpn.} \textbf{\bibinfo{volume}{85}},
  \bibinfo{pages}{043706} (\bibinfo{year}{2016}).

\bibitem[{\citenamefont{Nakajima et~al.}(2016)\citenamefont{Nakajima, Tomita,
  Taie, Ichinose, Ozawa, Wang, Troyer, and Takahashi}}]{Nakajima:2016aa}
\bibinfo{author}{\bibfnamefont{S.}~\bibnamefont{Nakajima}},
  \bibinfo{author}{\bibfnamefont{T.}~\bibnamefont{Tomita}},
  \bibinfo{author}{\bibfnamefont{S.}~\bibnamefont{Taie}},
  \bibinfo{author}{\bibfnamefont{T.}~\bibnamefont{Ichinose}},
  \bibinfo{author}{\bibfnamefont{H.}~\bibnamefont{Ozawa}},
  \bibinfo{author}{\bibfnamefont{L.}~\bibnamefont{Wang}},
  \bibinfo{author}{\bibfnamefont{M.}~\bibnamefont{Troyer}}, \bibnamefont{and}
  \bibinfo{author}{\bibfnamefont{Y.}~\bibnamefont{Takahashi}},
  \bibinfo{journal}{Nat Phys} \textbf{\bibinfo{volume}{12}},
  \bibinfo{pages}{296} (\bibinfo{year}{2016}).

\bibitem[{\citenamefont{Lohse et~al.}(2016)\citenamefont{Lohse, Schweizer,
  Zilberberg, Aidelsburger, and Bloch}}]{Lohse:2016aa}
\bibinfo{author}{\bibfnamefont{M.}~\bibnamefont{Lohse}},
  \bibinfo{author}{\bibfnamefont{C.}~\bibnamefont{Schweizer}},
  \bibinfo{author}{\bibfnamefont{O.}~\bibnamefont{Zilberberg}},
  \bibinfo{author}{\bibfnamefont{M.}~\bibnamefont{Aidelsburger}},
  \bibnamefont{and} \bibinfo{author}{\bibfnamefont{I.}~\bibnamefont{Bloch}},
  \bibinfo{journal}{Nat Phys} \textbf{\bibinfo{volume}{12}},
  \bibinfo{pages}{350} (\bibinfo{year}{2016}).

\bibitem[{\citenamefont{Thouless}(1983)}]{Thouless:1983fk}
\bibinfo{author}{\bibfnamefont{D.~J.} \bibnamefont{Thouless}},
  \bibinfo{journal}{Phys. Rev. B} \textbf{\bibinfo{volume}{27}},
  \bibinfo{pages}{6083} (\bibinfo{year}{1983}).

\bibitem[{\citenamefont{Rice and Mele}(1982)}]{Rice:1982qf}
\bibinfo{author}{\bibfnamefont{M.~J.} \bibnamefont{Rice}} \bibnamefont{and}
  \bibinfo{author}{\bibfnamefont{E.~J.} \bibnamefont{Mele}},
  \bibinfo{journal}{Phys. Rev. Lett.} \textbf{\bibinfo{volume}{49}},
  \bibinfo{pages}{1455} (\bibinfo{year}{1982}).

\bibitem[{\citenamefont{Vanderbilt and King-Smith}(1993)}]{Vanderbilt:1993fk}
\bibinfo{author}{\bibfnamefont{D.}~\bibnamefont{Vanderbilt}} \bibnamefont{and}
  \bibinfo{author}{\bibfnamefont{R.~D.} \bibnamefont{King-Smith}},
  \bibinfo{journal}{Phys. Rev. B} \textbf{\bibinfo{volume}{48}},
  \bibinfo{pages}{4442} (\bibinfo{year}{1993}).

\bibitem[{\citenamefont{Xiao et~al.}(2010)\citenamefont{Xiao, Chang, and
  Niu}}]{Xiao:2010fk}
\bibinfo{author}{\bibfnamefont{D.}~\bibnamefont{Xiao}},
  \bibinfo{author}{\bibfnamefont{M.-C.} \bibnamefont{Chang}}, \bibnamefont{and}
  \bibinfo{author}{\bibfnamefont{Q.}~\bibnamefont{Niu}}, \bibinfo{journal}{Rev.
  Mod. Phys.} \textbf{\bibinfo{volume}{82}}, \bibinfo{pages}{1959}
  (\bibinfo{year}{2010}).

\bibitem[{\citenamefont{Zak}(1989)}]{Zak:1989fk}
\bibinfo{author}{\bibfnamefont{J.}~\bibnamefont{Zak}}, \bibinfo{journal}{Phys.
  Rev. Lett.} \textbf{\bibinfo{volume}{62}}, \bibinfo{pages}{2747}
  (\bibinfo{year}{1989}).

\bibitem[{\citenamefont{Marzari and Vanderbilt}(1997)}]{Marzari:1997aa}
\bibinfo{author}{\bibfnamefont{N.}~\bibnamefont{Marzari}} \bibnamefont{and}
  \bibinfo{author}{\bibfnamefont{D.}~\bibnamefont{Vanderbilt}},
  \bibinfo{journal}{Phys. Rev. B} \textbf{\bibinfo{volume}{56}},
  \bibinfo{pages}{12847} (\bibinfo{year}{1997}).

\bibitem[{\citenamefont{Peschel}(2003)}]{Peschel:2003uq}
\bibinfo{author}{\bibfnamefont{I.}~\bibnamefont{Peschel}}, \bibinfo{journal}{J.
  Phys. A: Math.Gen.} \textbf{\bibinfo{volume}{36}}, \bibinfo{pages}{L205}
  (\bibinfo{year}{2003}).

\bibitem[{\citenamefont{Sheng et~al.}(2006)\citenamefont{Sheng, Weng, Sheng,
  and Haldane}}]{PhysRevLett.97.036808}
\bibinfo{author}{\bibfnamefont{D.~N.} \bibnamefont{Sheng}},
  \bibinfo{author}{\bibfnamefont{Z.~Y.} \bibnamefont{Weng}},
  \bibinfo{author}{\bibfnamefont{L.}~\bibnamefont{Sheng}}, \bibnamefont{and}
  \bibinfo{author}{\bibfnamefont{F.~D.~M.} \bibnamefont{Haldane}},
  \bibinfo{journal}{Phys. Rev. Lett.} \textbf{\bibinfo{volume}{97}},
  \bibinfo{pages}{036808} (\bibinfo{year}{2006}).

\bibitem[{\citenamefont{Fukui and Hatsugai}(2007)}]{Fukui:2007sf}
\bibinfo{author}{\bibfnamefont{T.}~\bibnamefont{Fukui}} \bibnamefont{and}
  \bibinfo{author}{\bibfnamefont{Y.}~\bibnamefont{Hatsugai}},
  \bibinfo{journal}{Phys. Rev. B} \textbf{\bibinfo{volume}{75}},
  \bibinfo{pages}{121403} (\bibinfo{year}{2007}).

\bibitem[{\citenamefont{Fukui et~al.}(2005)\citenamefont{Fukui, Hatsugai, and
  Suzuki}}]{FHS05}
\bibinfo{author}{\bibfnamefont{T.}~\bibnamefont{Fukui}},
  \bibinfo{author}{\bibfnamefont{Y.}~\bibnamefont{Hatsugai}}, \bibnamefont{and}
  \bibinfo{author}{\bibfnamefont{H.}~\bibnamefont{Suzuki}},
  \bibinfo{journal}{J. Phys. Soc. Jpn.} \textbf{\bibinfo{volume}{74}},
  \bibinfo{pages}{1674} (\bibinfo{year}{2005}).

\bibitem[{\citenamefont{Wang et~al.}(2013)\citenamefont{Wang, Troyer, and
  Dai}}]{Wang:2013fk_pump}
\bibinfo{author}{\bibfnamefont{L.}~\bibnamefont{Wang}},
  \bibinfo{author}{\bibfnamefont{M.}~\bibnamefont{Troyer}}, \bibnamefont{and}
  \bibinfo{author}{\bibfnamefont{X.}~\bibnamefont{Dai}},
  \bibinfo{journal}{Phys. Rev. Lett.} \textbf{\bibinfo{volume}{111}},
  \bibinfo{pages}{026802} (\bibinfo{year}{2013}).

\bibitem[{\citenamefont{Hatsugai and Fukui}(2016)}]{Hatsugai:2016aa}
\bibinfo{author}{\bibfnamefont{Y.}~\bibnamefont{Hatsugai}} \bibnamefont{and}
  \bibinfo{author}{\bibfnamefont{T.}~\bibnamefont{Fukui}},
  \bibinfo{journal}{Phys. Rev. B} \textbf{\bibinfo{volume}{94}},
  \bibinfo{pages}{041102} (\bibinfo{year}{2016}).

\end{thebibliography}

\end{document}